 \documentstyle[prl,aps]{revtex}
\newcommand{\ket}[1]{{|#1\rangle}}
\newcommand{\bra}[1]{{\langle#1|}}
\newcommand{\binomial}[2]{{{#1}\choose{#2}}}

\begin{document}

\preprint{28-August-96 RL DRAFT-1}

\draft

\twocolumn
\title{Quantum MacWilliams Identities}

\author{Peter Shor$^{1,3}$
and Raymond Laflamme$^{2,3}$}
 
\address{\vspace*{1.2ex} 
         \hspace*{0.5ex}${}^1$AT\&T Research, Room 2D-149, 600 Mountain Ave. \\
         Murray Hill, NJ 07974, USA \\
 	\hspace*{0.5ex}{${}^2$Theoretical Astrophysics, T-6, MS B288}\\
 	 Los Alamos National Laboratory, Los Alamos, NM 87545, USA\\
 	\hspace*{0.5ex}{${}^3$Institute for Theoretical Physics}\\
 	 University of California, Santa Barbara, CA 93106-4030, USA}

\date{\today}
\maketitle

%%%%%%%%%%%%%%%%%%%%%%%%%%%%%%%%%%%%%%%%%%%%%%%%%%%%%%%%%%%%%%%%%%%%%%%%%%%%%
% Abstract

\begin{abstract}

We derive a relationship between two different notions of fidelity
(entanglement fidelity and average fidelity)
for a completely depolarizing quantum channel.  
This relationship gives rise to a quantum analog of the MacWilliams
identities in classical coding theory.  These identities
relate the weight enumerator of a code to the one of its dual
and, with linear programming techniques, provided a powerful tool
to investigate the possible existence of codes.  
The same techniques can be adapted
to the quantum case. We give examples of their power.

\end{abstract}


% \narrowtext

%%%%%%%%%%%%%%%%%%%%%%%%%%%%%%%%%%%%%%%%%%%%%%%%%%%%%%%%%%%%%%%%%%%%%%%%%%%%%

%Main Text

The discovery of error correcting codes\cite{shor95,steaneprl} for quantum computers
has revolutionized the field of quantum information.  Although quantum
computing holds great promise, it is plagued by the fragility of 
quantum information\cite{Landauer95,Unruh94,CLSZ95}.
Quantum error correction is a technique 
which enables one to encode quantum information in a robust way
and therefore overcome this fragility. 

It is important to classify codes in order to know
what is the most compact way to encode a number of qubits against a
given number of errors.
Various techniques have been used to discover such codes\cite{shor95,calde,steane95,LMPZ,manlaf,bennett,gott,steane96a,calde96,steane96b}.
In the classical theory a very powerful technique for looking for 
the existence of codes is the use of MacWilliams identities\cite{macsloane}.  
They relate the weight distribution of a code to the weight distribution
of its dual code.  These relationships can be used with linear
programming to find bounds on how good quantum codes can be and to test
whether potential good codes can exist \cite{McEliece}.

In this letter we give a quantum analog to the
weight distributions  which obey
the (classical) MacWilliams identities.  We use these identities to
derive the non-existence of some codes.  For example we will show that there
is no degenerate 5 bit code which encodes 1 qubit of information and
corrects for a general 1 bit error; this implies that
the perfect code of \cite{LMPZ,bennett} is the best that can attained 
in this respect.  We will also show that there is no 9-bit code
which encodes 1 qubit of information and corrects for a general 2 qubit error.

Let us first introduce a basis of operators given by the Hermitian set  
\begin{equation}
{\cal E} =  \{ \sigma^1_i \otimes \sigma^2_j ... \otimes \sigma^n_k \}
\end{equation}
for all $i,j, ... k$ and 
where $\sigma^n_k$ is chosen from
the set of Pauli matrices augmented by the identity,
acting on the $n^{th}$ qubit.  
Note that all elements of ${\cal E}$ give the identity when multiplied 
by their Hermitian conjugate.
We define the set $E_d$ as the subset of  ${\cal E}$ containing exactly $d$ Pauli 
matrices different from the identity, and we call it the set of distance $d$.

In a system interacting with an environment, errors (differences
from the original state) can be classified
as bit flip, sign flip or bit and sign flip\cite{shor95,LMPZ}
corresponding to the three Pauli matrices.
The set ${\cal E}$ corresponds to all possible effects due to 
independent environments. If we assigned an equal probability of $(1-p)/3$
every time a Pauli matrix appears in a member of ${\cal E}$
an initial state $\rho_i$ would therefore evolve as 
\begin{equation}
\rho_f = \sum_{{E \in \cal E}} p^{d(E)} (\frac{1-p}{3})^{n-d(E)} E\rho_i E^\dagger
\end{equation}
where  $\rho_f$ is the final state
and $d(E)$ is the distance of the operator $E$.

We now define two weights $A_d$ and $B_d$ on operators 
${\cal O}_1,{\cal O}_2$, where $d$ ranges from 0 to $n$,
as
\begin{equation}
A_d = \frac{1}{{\rm tr}{\cal O}_1 {\rm tr}{\cal O}_2}
\sum_{E_d} {\rm tr} (E_d{\cal O}_1) {\rm tr} (E^\dagger_d {\cal O}_2)
\end{equation}
\begin{equation}
B_d = \frac{1}{{\rm tr}{\cal O}_1{\cal O}_2}
\sum_{E_d} {\rm tr} (E_d {\cal O}_1 E^\dagger_d {\cal O}_2)
\end{equation}
where the sum is over all $E_d$ of distance $d$.

We define the weight enumerator as
\begin{equation}
A(z) = \sum_{d=0}^{n} A_d z^d \nonumber 
\end{equation}
and a similar equation for $B$.  
The MacWilliams identities are
relationships between the weight enumerators $A(z)$ and $B(z)$
given by
\begin{equation}
B(z) = \frac{{\rm tr}{\cal O}_1 {\rm tr}{\cal O}_2}
            {2^{n}{\rm tr}{\cal O}_1{\cal O}_2} 
(1+3z)^n A(\frac{1-z}{1+3z})
\label{macw}
\end{equation}
[These are MacWilliams identities for codes over for GF(4)].
The proof uses the expansion ${\cal O}_i$ in term of the set ${\cal E}$ as
\begin{equation}
{\cal O}_i = \sum_{D\in {\cal E}} \frac{{\rm tr}(D^\dagger {\cal O}_i)}{2^n} D
\end{equation}
We can rewrite $B_d$ as
\begin{equation}
B_d =\frac{1}{{{\rm tr}{\cal O}_1{\cal O}_2}} 
\sum_{D,E_d,D'} {\rm tr}(E_d D  {E^\dagger}_d {D'} )
\frac{{\rm tr} (D^\dagger{\cal O}_1)}{2^n}
\frac{{\rm tr} ({D'}^\dagger {\cal O}_2)}{2^n}
\end{equation}
It is easy to convince yourself that we must have
$D  = D'$ for the trace ${\rm tr}(E_d D  {E^\dagger}_d {D'})$
to be non-zero as otherwise there would be 
a Pauli matrix operating on at least one qubit.
We can now see how to relate $B_d$ to a sum of $A_{d'}$ by deriving
the coefficient for every $D$ of weight $d'$ which is equal to
\begin{equation}
\alpha_{dd'} =\frac{{\rm tr}({\cal O}_1 {\cal O}_2)}{2^{2n}}
\sum_{E_d}  {\rm tr}(E_d D  {E^\dagger}_d D )
\label{alddp}
\end{equation}
for $D \in {\cal E}$. 
To prove the relationship
we need  to prove it for only one element $D$ of distance $d'$ as all
the others can be reached by permutations of the qubits 
and transformations which are tensor products of 1-qubit unitary
transformations.  Eqs.~(3-4) are invariant under these transformations.

We will be interested in the case where ${\cal O}_1= {\cal O}_2=P_c$, 
where $P_c$ is a projection operator defining a quantum code.
For 2 qubits the coefficient are given by
\begin{eqnarray}
\alpha_{00} = 2^n  \ \ \ \ \ ; \ 
\alpha_{01} = 2^n \ \ \ \ \ \  \ \ ; \
\alpha_{02}  = 2^n  \nonumber\\
\alpha_{10}  = 6*2^n  \ ; \
\alpha_{11} = -2*2^n \ ; \
\alpha_{12} = 2^n  \nonumber \\
\alpha_{20} = 9*2^n \ ; \
\alpha_{21} = -3*2^n \ ; \
\alpha_{22} = 2^n  
\end{eqnarray}
from which we deduce the relationship
\begin{eqnarray}
B_0 &=& \frac{1}{2^{2-k}} (A_0 + A_1 + A_2) \\
B_1 &=& \frac{1}{2^{2-k}} (6A_0 + 2A_1 - 2A_2) \\
B_2 &=& \frac{1}{2^{2-k}} (9A_0 - 3A_1 + A_2) 
\end{eqnarray} 
where ${\rm tr} (P_c)= 2^k$.

In general, 
\begin{equation}
\alpha_{dd'} = 2^n\sum_{s=0}^d (-1)^s 3^{d-s} \binomial{d'}{s} 
\binomial{n-d'}{d-s},
\label{kraw}
\end{equation}
where the $s$'th term in the sum comes from considering 
the case in Eq(\ref{alddp}) where there are exactly $s$ qubits on which 
Pauli matrices act in $E_d$ and in $D$ simultaneously.
Eq.~(\ref{kraw}) is the standard expansion of the MacWilliams identity
in terms of Krawtchouk polynomials \cite{macsloane} and the 
MacWilliams identity (\ref{macw}) then follows from this expansion.

The origin of this relationship can be traced back to the fact that
the matrix $H_{ij}={\rm tr}E_iE_jE_i E_j$ is proportional to a 
Hadamard matrix. As in the classical case, it is a ``coarse grained'' version
of $H_{ij}$ which enters Eq.(\ref{alddp}).

In the case where $P_c$ is a projection operators in the subspace defined
by the set of states $\{ c_i\}$ we can rewrite the weights as

\begin{equation}
A_d = \frac{1}{2^{2k}}\sum_{E_d} \big|\sum_i \langle c_i|E_d|c_i\rangle \big|^2
\label{adcode}
\end{equation}

\begin{equation}
B_d = \frac{1}{2^{k}}\sum_{E_d} \sum_{ij} |\langle c_i|E_d|c_j\rangle |^2
\label{bdcode}
\end{equation}

>From the Cauchy-Schwartz inequality we deduce that these are 
non-negative numbers with $B_d\geq A_d$.
This is because the $B$'s are defined as a sum of the modulus squared
of every element of the operators of weight $d$ projected on the code
while the $A$'s are the squared modulus of a sum.

For a depolarizing channel with probability of distance 1 error
being $(1-p)$, the weight enumerator $A$ has the physical interpretation
that $p^n A((1-p)/3p)$   is the
fidelity of entanglement\cite{nielsen96}.  This is the probability
that a completely entangled state constructed from the 
basis states of the code remains intact after going through the channel.
The physical interpretation of $B$ is that
$p^n B((1-p)/3p)/{\rm tr}(P_c)$ is the average fidelity \cite{nielsen96}, 
i.e., the 
average probability over the states of an incoherehnt ensemble given by $P_c$ 
going through the channel and giving the same states.  

Necessary and sufficient condition for the quantum code ${\cal C}$ 
to correct $\lfloor d/2\rfloor$ errors are \cite{manlaf} that 
for all basis elements $\ket{c_a}$, $\ket{c_b}$ ($a\not= b$) of $P_c$ 
\begin{eqnarray}
\bra{c_a}E_{d'} \ket{c_a} &=& \bra{c_b} E_{d'} \ket{c_b}
\label{eqn:char1}
\end{eqnarray}
and
\begin{eqnarray}
\bra{c_a}E_{d'} \ket{c_b} &=& 0.
\label{eqn:char2}
\end{eqnarray}
for all elements $ E_{d'}$ of distance less or equal to $d$.
For a degenerate code [i.e. when (\ref{eqn:char1}) is non-zero],
we can deduce from Eqs.~(\ref{adcode},\ref{bdcode}) that 
$A_{d'} = B_{d'}$ for $ 1 \leq  d' \leq d$,
and these quantities are zero for a non-degenerate code.
Thus the property of error correction restricts the possible form of the weights.
The existence of non-negative weights is a necessary condition for 
a quantum error correcting code to exist.

As a first example of the power of these inequalities,
we look for the possible existence of a degenerate 5-bit code
which protects 1 qubit of information against a 
general 1 qubit error. This implies we are looking for
a code with $n=5$ and $k=2$ which satisfies the equations and inequalities
\begin{eqnarray}
 B_0 =  \frac{A_0 + A_1 + A_2 + A_3 + A_4 + A_5}{16} = A_0=1 \nonumber \\
 B_1 = \frac{15 A_0 + 11A_1 + 7A_2 + 3A_3 - A_4 - 5A_5}{16} = A_1   \nonumber \\
 B_2 = \frac{45 A_0 + 21A_1 + 5A_2 - 3A_3 -3A_4 + 5A_5}{8} = A_2  \nonumber \\
 B_3 = \frac{135 A_0 + 27A_1 - 9A_2 - 5A_3 + 7A_4 - 5A_5}{8} \geq A_3 \nonumber \\
 B_4 = \frac{405 A_0 - 27A_1 - 27A_2 + 21A_3 - 11A_4 + 5A_5}{16} \geq A_4 \nonumber \\
 B_5 = \frac{243 A_0 - 81A_1 + 27A_2 - 9A_3 + 3A_4 - A_5}{16} \geq A_5. 
\nonumber
\end{eqnarray}

This is a set of linear equations and inequalities
in the $A_i$, which can easily be solved using linear programming techniques.
We find that the only solution is given by
$A_i=(1,0,0,0,15,0)$ and therefore $B_i=(1,0,0,30,15,18)$.
This is the unique solution and since $A_1 = A_2 = 0$
it corresponds to a non-degenerate code.  Thus  no  degenerate code 
exists for 5 bits.  An explicit code with this weight enumerator 
was found in \cite{LMPZ,bennett}.

In a similar way, we can also show that it is not possible to find a code
which protects 1 qubit of information against 2 errors using $n=9$ qubits.
A solution of the Macwilliams identities exists for codes mapping 1 qubit
into $n=10$ qubits; however an extension of the techniques in this paper 
based on classical shadow code techniques rules this possibility out 
as well \cite{rains}.  
The smallest possible code protecting against two errors thus would map 
1 qubit into $n=11$ qubits; such a code was constructed in \cite{calde96}.

Both possibilities eliminated above would have required degenerate quantum codes. 
These might have allowed to find more compact codes
than would have been expected from an analogy to classical codes.
A systematic study of the MacWilliams identities for $n\leq 30$ \cite{rains} 
shows that this is not the case.  The most compact codes appear not to be
degenerate.  It will be interesting to know if this holds as $n\rightarrow\infty$.

In conclusion, we have derived the quantum analog of 
the MacWilliams identities which give necessary conditions 
for the existence of codes.
We have demonstrated the power of these identities by
showing the non-existence of certain degenerate codes using linear 
programming techniques.  The quantum Macwilliams identities will lead to a strong bound
on the existence of quantum codes as the number of qubits
grows large.  This will be important to understand the capacity of noisy 
quantum channels\cite{lloyd,nielsen96}
  
We would like to thank E. Knill for useful comments.
We are also grateful to D. DiVincenzo and W. Zurek
for the invitation to participate to the
the Quantum Coherence and Decoherence workshop
in Santa-Barbara.
This research was supported in part by the National Science Foundation under
Grant No. PHY94-07194.

\end{document}